\documentclass[twocolumn,aps,prc,superscriptaddress,showpacs,floatfix]{revtex4}
\usepackage{amssymb}
\usepackage{amsmath}
\usepackage{graphicx}

\usepackage[normalem]{ulem}
\usepackage[dvips]{color}
\usepackage{leftidx}
\usepackage{mhchem} % \ce{^{227}_{90}Th}
\renewcommand\sout{\bgroup \color{red} \ULdepth=-.5ex \ULset}

\begin{document}

\title{Production of $\Lambda\Lambda$ and $\overline{\Lambda \text{n}}$ in central Pb+Pb collisions at $\sqrt{s_{NN}}$=2.76  TeV
within a covariant coalescence model}
\author{Kai-Jia Sun}
\affiliation{Department of Physics and Astronomy and Shanghai Key Laboratory for Particle
Physics and Cosmology, Shanghai Jiao Tong University, Shanghai 200240, China}
\author{Lie-Wen Chen\footnote{%
Corresponding author: lwchen$@$sjtu.edu.cn}}
\affiliation{Department of Physics and Astronomy and Shanghai Key Laboratory for Particle
Physics and Cosmology, Shanghai Jiao Tong University, Shanghai 200240, China}
\affiliation{Center of Theoretical Nuclear Physics, National Laboratory of Heavy Ion
Accelerator, Lanzhou 730000, China}
\date{\today}

\begin{abstract}
We study the production of $\Lambda\Lambda$ and $\overline{\Lambda \text{n}}$
exotic states in central Pb+Pb collisions at $\sqrt{s_{NN}}=2.76$ TeV at LHC via both
hadron and quark coalescence within a covariant coalescence model with a
blast-wave-like parametrization for the phase-space configurations of constituent
particles at freezeout. In the hadron coalescence,
the two states are considered as molecular states while they are considered as
six-quark states in the quark coalescence.
For $\overline{\Lambda \text{n}}$,
we find that the yields of both molecular and six-quark states are much larger
than the experimental upper-limits.
For $\Lambda\Lambda$, while the molecule-state yield is much larger than the
experimental upper-limits, the six-quark-state yield could be lower than the
upper-limits.
The higher molecule-state yields
are mainly due to the large contribution of short-lived strong resonance decays into
(anti-)nucleons and (anti-)$\Lambda$ which can significantly enhance the molecule-state 
yields of $\Lambda\Lambda$ and $\overline{\Lambda \text{n}}$ via hadron coalescence. 
Our results suggest that the current experimental measurement 
at LHC cannot exclude the existence of the $\Lambda\Lambda$ as an exotic six-quark 
state, and if $\Lambda\Lambda$ is a six-quark state, it is then on the brink of being 
discovered.

\end{abstract}

\pacs{25.75.-q, 25.75.Dw}
\maketitle

\section{Introduction}
\label{introduction}
The conventional hadron spectrum can be well understood in
quark model in which mesons are considered to consist of a quark and 
an anti-quark while (anti)baryons consist of three (anti)quarks.
However, the fundamental theory of strong interaction,
Quantum Chromodynamics (QCD) does not forbid the existence of exotic
states with more than three valence quarks such as tetraquarks, pentaquarks,
hexquarks, and so on.
Jaffe~\cite{Jaf77} first predicted the existence of H-dibaryon,
a hypothetical bound state consisting of $uuddss$ with spin-parity $J^\pi=0^+$,
using a bag model approach.
Lattice QCD (LQCD) calculations~\cite{Bea11,Ino11}
also suggest the bound H-dibaryon with strangeness $S=-2$ and spin-parity
$J^\pi=0^+$ may exist, although this needs to be confirmed by the physical
point simulations~\cite{Doi15}.
A relevant and interesting question is about the possible existence of other exotic
dibaryon states including nucleon-hyperon bound states like $\Lambda n$
with $S=-1$ and $J^\pi=1^+$~\cite{Ste12}.
These studies would be extremely useful for understanding the largely uncertain
nucleon-hyperon and hyperon-hyperon interactions.

To understand the $\Lambda$-$\Lambda$ interaction,
the STAR collaboration~\cite{Ada15} at Relativistic Heavy-Ion Collider (RHIC)
recently measured the $\Lambda$-$\Lambda$ correlation
function for centrality $0-80\%$ Au+Au collisions at
$\sqrt{s_{NN}}=200$~GeV.
In Ref.~\cite{Ada15}, a positive scattering length $a_0>0$
(decreasing phase shift) is obtained by analyzing the data using
Lednicky-Lyuboshits~(LL) model formula. However, an opposite sign
of $a_0$ is favored~\cite{Mor15} by using Koonin-Pratt~(KP) formula.
These two apparently contradictive conclusions can be understood~\cite{Ohn16}
in the LL model formula with different values of $\Lambda$-$\Lambda$
pair purity probability $\lambda$ which is still largely uncertain.
Therefore, more experimental measurements are needed
to further constrain the details of the $\Lambda$-$\Lambda$ interaction
so as to determine if the bound $\Lambda \Lambda$ state can exist or not.

Very recently,
the ALICE collaboration~\cite{Ada16} at Large
Hadron Collider (LHC) searched for weakly decaying
$\Lambda\Lambda$ and $\overline{\Lambda n}$ exotic bound states
in central Pb+Pb collisions at $\sqrt{s_{NN}}=2.76$~TeV, and
no bound states of $\Lambda\Lambda$ and $\overline{\Lambda n}$
are measured in the decay modes $\overline{\Lambda n}\rightarrow \bar{d}\pi^+$
and H-dibaryon $\rightarrow \Lambda p\pi^-$.
The obtained upper-limits of the yields
($dN/dy$ at mid-rapidity) are found to be, in a large range of decay branching
ratio (BR), much lower than the predictions of various models, including the
equilibrium thermal model~\cite{sta14}, the non-equilibrium thermal model~\cite{Tor05,Tor06},
a hybrid UrQMD calculation~\cite{Ste12}, and the coalescence model in which 
they are assumed to be molecular states (with hadron coalescence) or 
six-quark states (with quark coalescence)~\cite{Cho11,Cho11prc}.
The comparison between
the experimental measurement and these existing model predictions seems to
exclude the existence of the $\Lambda\Lambda$ and
$\overline{\Lambda n}$ exotic bound states.
However, we note that the coalescence model predictions of
H-dibaryon in Refs.~\cite{Cho11,Cho11prc}) are made only for central Au+Au (Pb+Pb)
collisions at $\sqrt{s_{NN}}=200$ GeV ($5.5$ TeV) at RHIC (LHC) based on a significantly
simplified approximate analytic formula of the
coalescence model~\cite{Cho11,Cho11prc}.
Therefore, it is interesting to explore the production of $\Lambda\Lambda$ and
$\overline{\Lambda n}$ in central Pb+Pb collisions at
$\sqrt{s_{NN}}=2.76$~TeV in a more realistic coalescence model, and see if the 
experimental limits can or not exclude the existence of $\Lambda\Lambda$ and
$\overline{\Lambda n}$ as molecular states or six-quark states. 
This is the main motivation of the present work.

In this paper, we carefully calculate the yields of the
$\Lambda\Lambda$ and $\overline{\Lambda n}$
exotic states in central Pb+Pb collisions at $\sqrt{s_{NN}}=2.76$~TeV
by combining the covariant coalescence model~\cite{Dov91} together with a blast-wave-like
parametrization~\cite{Ret04} for the phase space configurations of
constituent particles at freezeout.
We demonstrate that while the yields of molecule-state and six-quark-state
$\overline{\Lambda n}$ as well as molecule-state $\Lambda\Lambda$
are much higher than the experimental upper-limits, the six-quark-state yield
of $\Lambda\Lambda$ could be lower than the experimental upper-limits.
Therefore, our results suggest that the current experimental measurement at LHC 
cannot exclude the existence of the $\Lambda\Lambda$ as an exotic six-quark state
although the $\Lambda\Lambda$ is unlikely to be a molecular state.

\section{Covariant Coalescence Model}

The coalescence model~\cite{Dov91,But61,Sat81,Cse86,Oh09,Xue15} and the
thermal model~\cite{Cle91,Bra95,Cle99,And11,Cle11,Ste12} provide two main approaches
to describe the composite particle production in relativistic heavy-ion collisions.
The main feature of
the coalescence model~\cite{But61,Sat81,Cse86} is that the coalescence probability
depends on the details of the phase space configurations of the constituent particles at
freezeout as well as the statistical weight and wave function
of the coalesced cluster, while these details are of no relevance in the thermal
model~\cite{Cle91,Bra95,Cle99} of cluster creation.
For particle production at mid-rapidity in central Pb+Pb collisions at
$\sqrt{s_{NN}}=2.76$~TeV  considered here,
we assume a longitudinal boost-invariant expansion for the constituent
particles and
the Lorentz invariant one-particle momentum distribution is then given by
\begin{eqnarray}
E\frac{d^3N}{d^3p}=\frac{d^3N}{p_Tdp_T d\phi_p dy } =  \int d^4x S(x,p),
\end{eqnarray}
where
$S(x,p)$ is the emission function which
is taken to be a blast-wave-like parametrization as~\cite{Ret04}
\begin{eqnarray}
S(x,p)d^4x = m_T cosh(\eta-y)f(x,p)J(\tau)d\tau d\eta rdrd\phi_s,
\end{eqnarray}
where $m_T=\sqrt{m^2+p_T^2}$ is the transverse mass of the emitted
particle and $f(x,p)$ is the statistical distribution function
which is given
by $f(x,p)=g(2\pi)^{-3}[\exp(p^{\mu}u_{\mu}/kT)/\xi \pm 1]^{-1}$
with $g$ being spin degeneracy factor, $\xi$ the fugacity, $u_{\mu}$ the
four-velocity of a fluid element in the fireball, and $T$ the local temperature.
The freezeout time is assumed to follow a
Gaussian distribution
$J(\tau)=\frac{1}{\Delta \tau \sqrt{2\pi}}\exp(-\frac{(\tau-\tau_0)^2}{2(\Delta \tau)^2})$
with a mean value $\tau_0$ and a dispersion $\Delta \tau$.
The transverse rapidity distribution of the fluid element in the fireball is
parametrized as $\rho=\rho_0 r/R_0$ with $\rho_0$ being the maximum transverse
rapidity and $R_0$ the transverse radius of the fireball~\cite{Ret04}.
The detailed information can be found in Ref.~\cite{SunKJ15}.
In the present coalescence model, the phase space
freezeout configuration of constituent particles
are thus determined by six parameters, i.e., $T$, $\rho_0$, $R_0$, $\tau_0$,
$\Delta \tau$ and $\xi$.

In the coalescence model, the cluster production probability can
be calculated through the overlap of the cluster Wigner function with the
constituent particle phase-space distribution at freezeout. The invariant
momentum distribution of the formed cluster consisting of $M$
constituent particles can be obtained as
\begin{eqnarray}
E\frac{d^3N_c}{d^3P}&=&Eg_c\int  \bigg(\prod_{i=1}^{M} \frac{d^3p_i }{E_i}d^4x_iS(x_i,p_i)\bigg)\times \notag \\
&&\rho_c^W(x_1,...,x_M;p_1,...,p_M)\delta^3(\mathbf{P}-\sum_{i=1}^M\mathbf{p_i}),
\label{Eq:Coal}
\end{eqnarray}
where $N_c$ is the multiplicity of the cluster with energy (momentum) $E$ ($\mathbf{P}$),
$g_c$ is the coalescence factor, $\rho_c^W$ is the cluster Wigner function
and the $\delta$-function is adopted to ensure momentum conservation.
In this work, the harmonic oscillator wave functions are assumed for the clusters
in the rest framework and
the corresponding Wigner function is $\rho_c^W(x_1,...,x_M;p_1,...,p_M)
= \rho ^{W}(q_{1},\cdot \cdot \cdot ,q_{M-1},k_1,\cdot
\cdot \cdot ,k_{M-1})
= 8^{M-1}\exp [-\sum_{i=1}^{M-1}(q_{i}^{2}/\sigma _{i}^{2}+\sigma_{i}^{2}k_i^{2})]$,
where $\mu_{i-1}= \frac{i}{i-1} \frac{m_i \sum_{k=1}^{i-1}m_k}{\sum_{k=1}^{k=i}m_k},(i\geq2)$
is the reduced mass,
$\sigma_i^2 = (\mu_{i} w)^{-1}(1\leq i\leq M-1)$, and $w$ is the harmonic oscillator frequency.
The details about the coordinate transformation from
$(x_1,...,x_M)[(p_1,...,p_M)]$ to relative coordinates $(q_1,...,q_{M-1})[(k_1,...,k_{M-1})]$
can be found in Ref.~\cite{SunKJ15}.
The mean-square radius is given by
$<r^2_M> = \frac{3}{2M w}[\sum_{i=1}^M \frac{1}{m_i} -\frac{M}{\sum_{i=1}^M m_i}]$.
The integral~(\ref{Eq:Coal}) can be calculated directly through multi-dimensional
numerical integration by Monte-Carlo method~\cite{Lep78,SunKJ15}.
 Since
the constituent particles may have different freezeout time, in the numerical
calculations, the particles that freeze out earlier are allowed to propagate
freely until the time when the last particle in the cluster freezes out
in order to make the coalescence at equal time~\cite{Mat97,ChenLW06,SunKJ15}.

\section{result and discussion}

\begin{table}
\caption{Parameters of the blast-wave-like analytical parametrization for (anti-)nucleon~\cite{SunKJ15}, (anti-) $\Lambda$~\cite{SunKJ152} and light quark phase-space  configurations for Pb+Pb collisions at $\protect\sqrt{s_{NN}}=2.76${\protect\small \ TeV}. }
\begin{tabular}{cccccccc}
        \hline \hline
          & T(MeV) & $\rho_0$ & $R_0$(fm) & $\tau_0$(fm/c)& $\Delta \tau$  & $\xi_H$ \\
         \hline
        FOPb-N & 121.1  & 1.215 & 19.7 & 15.5& 1.0 & 3.72  \\
       % FOPb-$\Lambda$ & 121.1  & 1.215 & 19.7 & 15.5& 1.0 & 9.536   \\
       FOPb-$\Lambda^*$ & 123.4  & 1.171 & 16.7 & 13.1& 1.0 & 13.64   \\
        \hline  \hline
          & T(MeV) & $\rho_0$ & $R_0$(fm) & $\tau_0$(fm/c)& $\Delta \tau$  & $\xi_u$ & $\xi_s$  \\
         \hline
        FOPb-Q  & 154  & 1.08 & 13.6 & 11.0 & 1.3 & 1.02  &0.89  \\
        \hline  \hline
\end{tabular}
\label{ParamHadron}
\end{table}

\begin{figure}
\includegraphics[scale=0.33]{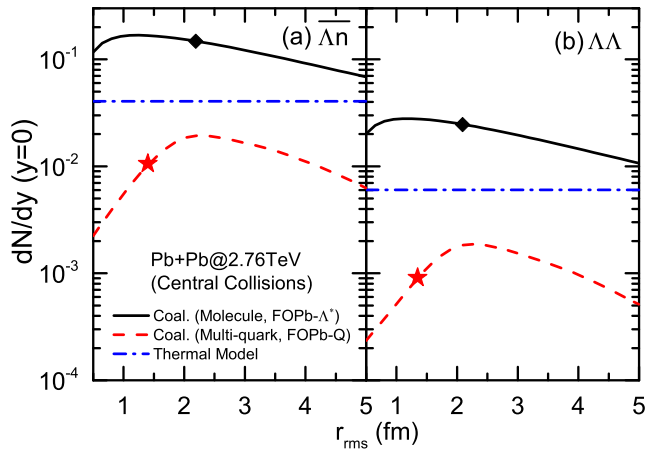}
\caption{The yields of $\overline{\Lambda n}$~(a) and
$\Lambda\Lambda$~(b) versus root-mean-square radii through hadron
coalescence and quark coalescence. The diamonds indicate the yield
with binding energy $E_{\text{bind}} = 1$ MeV in the
hadron coalescence while the stars represent the yield with a radius
obtained from the harmonic oscillator frequency $w_s$ (see text for
the details) in the quark coalescence.
The dash-dotted lines are predictions from the thermal model~\cite{Ada16}.}
\label{diLambHadron}
\end{figure}

\subsection{Hadron Coalescence}

Firstly, we consider $\Lambda\Lambda$ and $\overline{\Lambda n}$
as molecular states and their productions in central Pb+Pb collisions at
$\sqrt{s_{NN}}=2.76$~TeV at LHC can be described by $\Lambda$-$\Lambda$ and
$\overline{\Lambda}$-$\overline{\text{n}}$ coalescence, respectively, in the covariant
coalescence model. The basic inputs in the present coalescence model
calculations are the freezeout configurations of nucleons and $\Lambda$
particles which can be extracted from experimental information on the
production of protons, $\Lambda$, light nuclei and light hypernuclei.
For central Pb+Pb collisions at $\protect\sqrt{s_{NN} }=2.76$ TeV,
the freezeout configurations of nucleons (denoted as FOPb-N) and $\Lambda$
(denoted as FOPb-$\Lambda^*$) have been obtained by fitting the experimental
spectra of protons, $\Lambda$, deuterons and $^3$He as well as the measured
$^3_\Lambda \text{H}$/$^3\text{He}$ ratio, and the details
can be found in Refs.~\cite{SunKJ15,SunKJ152}.
The parameter values of FOPb-N and FOPb-$\Lambda^*$ are summarized
in Table~\ref{ParamHadron}, and one can see that
the protons and $\Lambda$ have different freezeout configurations
with the latter having an earlier mean freezeout time and a smaller
freezeout radius to reproduce the measured $^3_\Lambda \text{H}$/$^3\text{He}$
ratio and the $^3_\Lambda \text{H}$ spectrum, as discussed in detail
in Ref.~\cite{SunKJ152}.
In central Pb+Pb collisions at $\protect\sqrt{s_{NN} }=2.76$ TeV,
we assume the particles and their anti-partners
(as well as protons and neutrons) have the same freezeout
configuration.

In the coalescence model, the cluster yield also depends on the
cluster size. The root-mean-square radii~($r_{\text{rms}}$) of
$\Lambda\Lambda$ and $\overline{\Lambda n}$ can be related
to their binding energies~\cite{Cho11,Cho11prc} and their
values are treated as free parameters in this work.
Fig.~\ref{diLambHadron} shows the $p_T$-
integrated yield in the mid-rapidity region ($-0.5 \le y \le 0.5$)
(i.e., $dN$/$dy$ at $y=0$) of
$\overline{\Lambda n}$~(panel (a)) and $\Lambda\Lambda$~(panel (b))
versus $r_{\text{rms}}$ in the range of $0.5\sim 5$ fm through hadron
coalescence (solid lines).
It should be noted that the measured $\Lambda$ spectrum~\cite{Abe13}
includes the contributions from strong  and electro-magnetic decays
but excludes the contributions from weak
decays. Following Ref.~\cite{Cho11prc},
we assume the measured $\Lambda$ multiplicity
$N_{\Lambda(1115)}^{\text{measured}} = N_{\Lambda(1115)} + \frac{1}{3} N_{\Sigma(1192)} + (0.87+\frac{0.11}{3})N_{\Sigma(1385)} = 7.44N_{\Lambda(1115)}$ from which one can easily obtain that
the electro-magnetic decay contributes about $16.4\%$ of the
total yields. Because the $\Lambda$ particles coming from electro-magnetic
decay of $\Sigma(1192)$ do not appear in the fireball,
thus they should not contribute to the coalescence process and this has
been corrected for the results in Fig.~\ref{diLambHadron}
and the following analyses.
For FOPb-N and FOPb-$\Lambda^*$, the yield ($dN$/$dy$ at $y=0$) of
$\overline{\Lambda n}$ ranges from $6.66 \times 10^{-2}$
to $16.8 \times 10^{-2}$ and
the yield of $\Lambda\Lambda$ ranges from $1.03\times 10^{-2}$ to
$2.80\times 10^{-2}$. In experimental analysis~\cite{Ada16},
the binding energy $E_{\text {bind}}$ of both $\overline{\Lambda n}$
and $\Lambda\Lambda$ is assumed to be $1$ MeV,
and the corresponding
$r_{\text{rms}}$ of $\overline{\Lambda n}$ and $\Lambda\Lambda$ are
$2.2$ fm and $2.1$ fm~\cite{Cho11prc}, respectively, which are shown
by diamonds in the solid lines
in Fig.~\ref{diLambHadron}.
The corresponding yields at $E_{\text {bind}}=1$ MeV are $14.79\times 10^{-2}$
for $\overline{\Lambda n}$ and
$2.51\times 10^{-2}$ for $\Lambda\Lambda$.
The predictions of a thermal model with a temperature of $156$ MeV~\cite{Ada16}
are also included in Fig.~\ref{diLambHadron} (dash-dotted lines) for comparison.
From Fig.~\ref{diLambHadron},
one can see that the yields of both $\overline{\Lambda n}$
and $\Lambda\Lambda$ from the thermal model
can be a few times smaller than that of hadron
coalescence, depending on the size of $\overline{\Lambda n}$
and $\Lambda\Lambda$. This feature could be due to the earlier $\Lambda$ freezeout in the
coalescence model, which significantly enhance the yield of light
hypernuclei~\cite{SunKJ152}.

\subsection{Quark Coalescence}
% \begin{table} [!h]
%         \caption{Wigner function}
%         \centering
%        \begin{tabular}{c|c|c|c|c|c|c}
%          \hline \hline
%          % after \\: \hline or \cline{col1-col2} \cline{col3-col4} ...
%          property & $\omega$ & $\mu_1$ & $\mu_2$ & $\sigma_1$  & $\sigma_2$ & rms\\
%          p & 0.167  & 0.3  & 0.3 & 0.88 & 0.88& 0.88      \\
%           $\phi$&  0.167 & 0.5  &  & 0.68 &   & 0.59      \\
%           $\Omega$ &  0.167 & 0.5  & 0.5 & 0.68 &0.68 & 0.68 \\
%           $\Xi$ &  0.167 & 0.5  & 0.35 & 0.68 &0.82 &   0.76  \\
%           $\Lambda$ &  0.167 & 0.3  & 0.375 & 0.88 & 0.79& 0.83 \\
%          \hline \hline
%        \end{tabular}
%        \label{WigFun}
% \end{table}

Now we assume $\Lambda\Lambda$ and $\overline{\Lambda n}$
are six-quark states, i.e., $\Lambda\Lambda$ is a bound state of
$uuddss$ while $\overline{\Lambda n}$ is a bound state of
$\bar u$$\bar d$$\bar d$$\bar u$$\bar d$$\bar s$, and their yields can
then be calculated through the quark coalescence model.
The quark coalescence calculation needs the
information of phase-space distributions of $u, d, s$ quarks and
their anti-partners at freezeout, which are determined in the present
work by fitting the measured spectra of $p(uud)$, $\Lambda(uds)$,
$\Xi^-(dds)$, $\Omega^-(sss)$ and $\phi(s\bar{s})$ in the quark
coalescence model.
In the following, we assume $u$ and $d$ quarks have
the same mass of $300$ MeV while $s$-quark mass is $500$ MeV.
For protons, the mean-square charged radius is found to be
$<r^2_p> = 0.70706\pm 0.00066$ fm$^2$ from the recent measurement
of muon-atom ($\mu p$) Lamb shift~\cite{Oli14}, which leads to a
harmonic oscillator frequency ($w$) of $0.184$ GeV for the proton
wave function.
For $\Lambda$, $\Xi^-$, $\Omega^-$ and $\phi$, their radii are unknown
and generally depend on model predictions. For simplicity, following
Refs.~\cite{Cho11,Cho11prc}, we assume the harmonic oscillator wave
functions of these strange hadrons share the same frequency $w_s$.
For $\Omega^-$, LQCD simulation found
$\sqrt{<r^2_{\Omega^-}>}=0.573-0.596$ fm~\cite{Ale15}, while a recent
work based on a combined chiral and $1/N_C$ expansions method
found that $\sqrt{<r^2_{\Omega^-}>}=1.0$ fm~\cite{Flo15}
gives a better $\chi^2$ fit for mean-square charge radii of baryons.
We have tried both values, and  find that
$\sqrt{<r^2_{\Omega^-}>}=1.0$ fm gives a better fit of the measured
spectra of these strange hadrons and thus the corresponding frequency
$w_s$ is $0.078$ GeV.
Table~\ref{RmsHadron} summarizes the corresponding root-mean-square
radii of hadrons and the coalescence factors $g_c$ including
spin and color degree of freedom.

\begin{table}[!h]
         \caption{The root-mean-square radii of hadrons and  corresponding coalescence factor $g_c$, including the spin and color degree of freedom.}
         \centering
        \begin{tabular}{cccccc}
          \hline \hline
          % after \\: \hline or \cline{col1-col2} \cline{col3-col4} ...
          Hadron & $p$ & $\phi$ & $\Xi$ & $\Omega$  & $\Lambda$ \\ \hline
          $r_{\text{rms}}$ (fm) &  0.84 & 0.87  & 1.1 & 1.0 & 1.2 \\ \hline
          $g_c$ &  $\frac{2}{3^3\times 2^3}$ &$\frac{3}{3^2\times 2^2}$ & $\frac{2}{3^3\times 2^3}$ & $\frac{4}{3^3\times 2^3}$  &$\frac{2}{3^3\times 2^3}$ \\
          \hline \hline
        \end{tabular}
        \label{RmsHadron}
\end{table}

\begin{figure}
\includegraphics[scale=0.41]{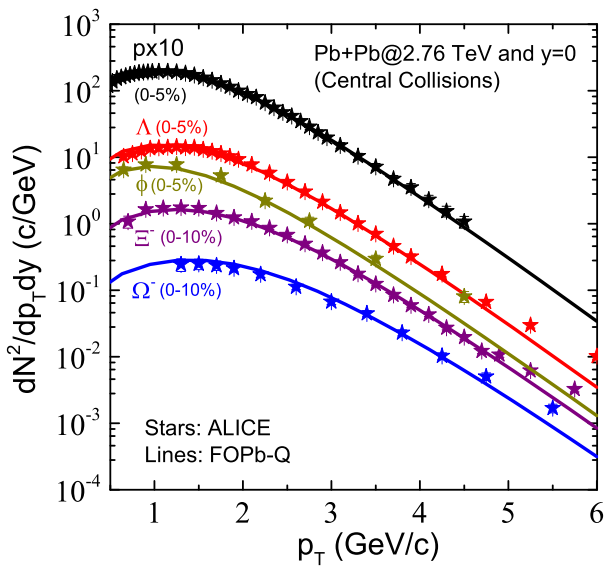}
\caption{Transverse momentum distributions of hadrons in central
Pb+Pb collisions at $\protect\sqrt{s_{NN}}=2.76 $ TeV. The experimental
results (stars) are taken from ALICE meassurement\cite{Abe12,Abe13,Abe15,Abe14}.
The solid lines are from the quark coalescence model.
The results of protons have been multiplied by a factor of $10$.}
\label{FigQuarkLHC}
\end{figure}

The weak decay contributions from heavier hadrons have been already excluded
in the experimental spectra of $p$, $\Lambda$, $\Xi^-$, $\Omega^-$ and
$\phi$, but not for the strong  and electro-magnetic decays.
In order to compare with experimental data, we have to include these effects
in quark coalescence calculations.
Following Refs.~\cite{Cho11,Cho11prc}, we assume the relations
$N_{\Lambda(1115)}^{\text{measured}} = N_{\Lambda(1115)} + \frac{1}{3} N_{\Sigma(1192)} + (0.87+\frac{0.11}{3})N_{\Sigma(1385)} = 7.44N_{\Lambda(1115)}$, $N_{p}^{\text{measured}} = N_{p} + N_{\Delta^{++}(1232)} + \frac{1}{2}N_{\Delta^{+}(1232)} + \frac{1}{2}N_{\Delta^{0}(1232)} = 5 N_p$, and $N_{\Xi^-}^{\text{measured}} = N_{\Xi^-} + \frac{1}{2}N_{\Xi(1530)} = 3 N_{\Xi^-}$.
For $\phi$ and $\Omega^-$, we assume no strong and electro-magnetic decay corrections, and thus
$N_{\phi}^{\text{measured}} = N_{\phi}$, $N_{\Omega^-}^{\text{measured}} = N_{\Omega^-}$.
In the above, $N_{\Lambda(1115)}$, $N_p$, $N_{\Xi^-}$, $N_{\phi}$ and $N_{\Omega^-}$
represent the corresponding hadron multiplicity obtained directly from the
quark coalescence model.

By fitting simultaneously the measured transverse momentum spectra
of $p$~\cite{Abe12}, $\Lambda$~\cite{Abe13}, $\phi$~\cite{Abe15}, $\Xi^-$~\cite{Abe14},
and $\Omega^-$~\cite{Abe14} for central Pb+Pb collisions
at $\protect\sqrt{s_{NN} }=2.76$ TeV, the parameters of the phase-space freezeout
configuration are extracted and summarized as FOPb-Q  in
Table~\ref{ParamHadron}. The temperature is fixed as $T = 154$ MeV~\cite{Baz12,Baz14}, and
the extracted transverse flow rapidity parameter is $\rho_0=1.08$, the transverse
radius is $R_0=13.6$ fm, the mean longitudinal proper freezeout time is
$\tau_0=11.0$ fm/c, the proper time dispersion is $\Delta \tau = 1.3$ fm/c,
and the fugacity is $1.02$ for $u$ and $d$ quarks and $0.89$ for $s$-quarks.
Fig.~\ref{FigQuarkLHC} compares the theoretical calculations with the
experimental data for the hadron transverse momentum
spectra.
In the theoretical calculations, as mentioned earlier,
the contributions from the strong and electro-magnetic decays
have been included to compare with the measured data.
One can see that the fit is very nice, and the obtained quark
freezeout configuration FOPb-Q can thus be used to predict
the corresponding yields of $\Lambda\Lambda$ and $\overline{\Lambda n}$
through the quark coalescence model.

The dashed lines in Fig.~\ref{diLambHadron} represent the $r_{\text{rms}}$
dependence of the yields of six-quark-state $\overline{\Lambda n}$
and $\Lambda\Lambda$ through quark coalescence using FOPb-Q.
The yield of $\overline{\Lambda n}$
ranges from $22.4\times 10^{-4}$ to $20.0\times 10^{-3}$
and correspondingly the yield of $\Lambda\Lambda$ ranges from
$2.35\times 10^{-4}$ to $1.95\times 10^{-3}$.
If we assume the harmonic oscillator frequencies for $\overline{\Lambda n}$
and $\Lambda\Lambda$ are the same as $w_s$, their $r_{\text{rms}}$ are
then found to be $1.4$ fm and $1.35$ fm, respectively, which
are indicated by stars in Fig.~\ref{diLambHadron} and the corresponding
yields are $10.6\times 10^{-3}$ and
$9.11\times 10^{-4}$, respectively.
It is seen that
the predicted yields by the quark coalescence are much smaller
than that of the hadron coalescence. This difference is
mainly due to the fact that in the hadron coalescence,
the strong decays dominantly contribute to the nucleon
and $\Lambda$ multiplicities, which will significantly enhance the yields of
$\overline{\Lambda n}$ and $\Lambda\Lambda$ through 
$\overline{\Lambda}$-$\overline{n}$ and $\Lambda$-$\Lambda$ 
coalescence, respectively.
In the quark coalescence, on the other hand, the six-quark-state
$\overline{\Lambda n}$ and $\Lambda\Lambda$ can only be produced
directly from the quark coalescence.

\subsection{Comparison with experimental limits}

\begin{figure}
\includegraphics[scale=0.315]{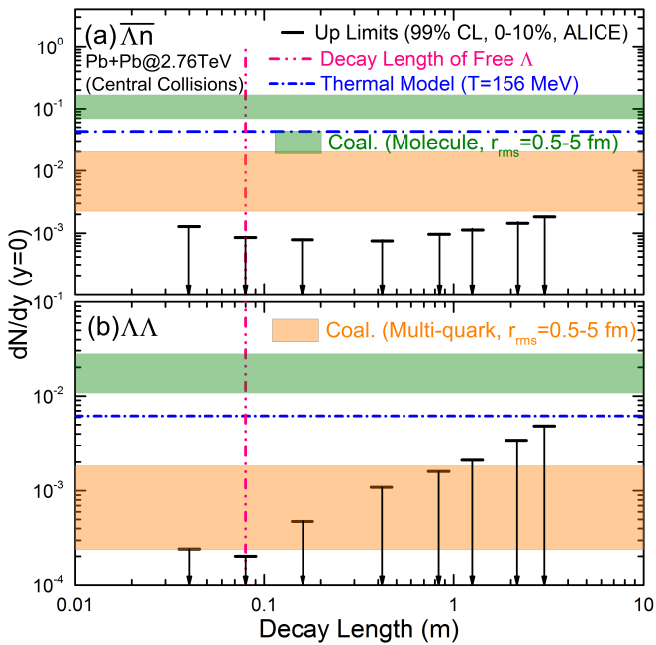}
\caption{Comparison between experimental upper-limits and theoretical
calculations for the yields of $\overline{\Lambda n}$~(a) and
$\Lambda\Lambda$~(b). A preferred branching ratio of $64\%$~\cite{sch00}
is used for $\Lambda\Lambda \rightarrow \Lambda p\pi^-$ and $54\%$~\cite{sch12}
for $\overline{\Lambda n}\rightarrow \bar{d}\pi^+$. The experimental
results are taken from ALICE measurement~\cite{Ada16}.
The olive shaded regions and the orange bands are predictions of molecular
states and multi-quark states from hadron coalescence and quark
coalescence, respectively.}
\label{Exp}
\end{figure}

Experimentally, no signals of bound states of $\overline{\Lambda n}$
and $\Lambda\Lambda$ are observed in central Pb+Pb collisions at
$\sqrt{s_{NN}}=2.76$ TeV, and instead the upper-limits of
these particle yields are obtained~\cite{Ada16}.
Shown in Fig.~\ref{Exp} is the comparison of
the yields ($dN$/$dy$ at $y=0$) of $\overline{\Lambda n}$~(panel (a))
and $\Lambda\Lambda$~(panel (b)) verse decay length (lifetime) between
experimental upper-limits and theoretical calculations, and here a preferred
branching ratio of $64\%$~\cite{sch00} is used for
$\Lambda\Lambda \rightarrow \Lambda p\pi^-$ and $54\%$~\cite{sch12}
for $\overline{\Lambda n}\rightarrow \bar{d}\pi^+$.
It is very interesting to see that
while the predicted yield of $\overline{\Lambda n}$ in either molecular
state or six-quark state is higher than the experimental upper-limit, the yield
of six-quark-state $\Lambda\Lambda$ could be lower than the experimental
upper-limit although the yield of molecule-state $\Lambda\Lambda$ is higher
than the upper-limit. These features indicate that the bound state
of $\overline{\Lambda n}$ is unlikely to exist, either in molecular
or in six-quark state, which is consistent with the conclusion obtained from
experimental analyses by the HypHI Collaboration~\cite{Rap13}. Meanwhile, the molecular state of $\Lambda\Lambda$ is
unlikely to exist either. However, the six-quark state of $\Lambda\Lambda$ could
exist, and cannot be excluded by the ALICE measurement.

\begin{figure}
\includegraphics[scale=0.325]{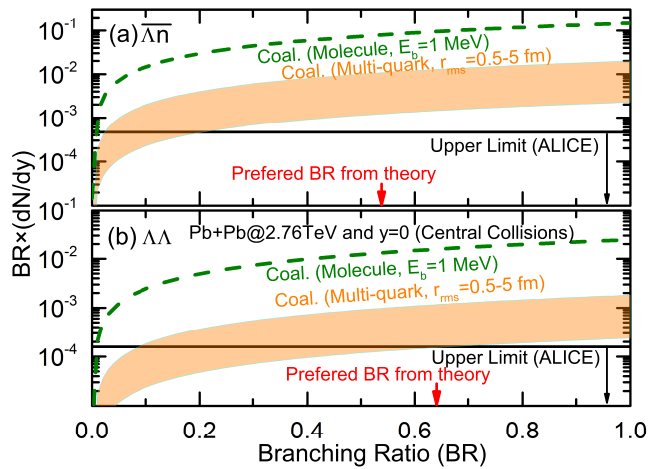}
\caption{Comparison between experiment upper-limits under the assumption
of the lifetime of a free $\Lambda$ and theoretical calculations for
the yields of $\overline{\Lambda n}$ (a) and $\Lambda\Lambda$ (b).
The dashed lines represent the molecular states from hadron
coalescence with a binding energy $E_{\text{bind}}=1$ MeV and the bands
correspond to multi-quark states from quark coalescence with
$r_{\text{rms}}=0.5\sim 5$ fm.}
\label{ExpBR}
\end{figure}

In Fig.~\ref{Exp}, the branching ratios of
$\Lambda\Lambda \rightarrow \Lambda p\pi^-$ and
$\overline{\Lambda n}\rightarrow \bar{d}\pi^+$ are fixed at their
preferred values which are obtained in model prediction.
To see the branching ratio dependence of our conclusion, we present in
Fig.~\ref{ExpBR} the comparison for the yields between experimental
upper-limits of $\overline{\Lambda n}$ (panel (a)) and
$\Lambda\Lambda$~(panel (b)) with a lifetime of  free $\Lambda$
and the corresponding theoretical calculations for different decay
branching ratios.
The dashed lines show the predictions from hadron coalescence with a
binding energy $E_{\text{b}} = 1$ MeV and the bands correspond
to the results from quark coalescence with radius from $0.5$ fm to
$5$ fm. One can see the yields of molecular states are much higher
than the upper-limits in almost the whole range of decay branching ratio.
The yield of six-quark-state $\overline{\Lambda n}$ can be
consistent with experimental upper-limits only when the decay branching
ratio is smaller than $0.2$ which is much smaller than the preferred value,
while the yield of six-quark-state $\Lambda\Lambda$ is consistent with
experimental upper-limits even assuming the decay branching ratio is as
large as $0.7$ which is larger than the preferred value.

\section{conclusion}

Based on a covariant coalescence model with a blast-wave-like
analytical parametrization for the phase-space configuration of
the constituent particles, we have elaborately studied the
productions of $\overline{\Lambda n}$ and $\Lambda\Lambda$
in central Pb+Pb collisions at $\sqrt{s_{NN}}=2.76$ TeV.
Their yields are calculated either through
hadron coalescence or quark coalescence. In the hadron coalescence,
the two states are considered as molecular states, and the freezeout
phase-space configurations of nucleons and $\Lambda$ are extracted
through fitting the measured spectra of light nuclei and hypernuclei.
In the quark coalescence, the two states are considered as
six-quark states, and the freezeout phase-space configuration of
light and strange quarks are extracted through fitting the measured
spectra of $p$, $\Lambda$, $\phi$, $\Xi^-$, and $\Omega^-$.

Our results have indicated that the yields of
$\overline{\Lambda n}$ and $\Lambda\Lambda$ in multi-quark states
from quark coalescence are much lower than that in molecular states
from hadron coalescence. This is mainly due to the dominant strong decay
contributions into nucleons and $\Lambda$ which can significantly enhance
the molecule-state yields through hadron coalescence.
In particular, we have found that although the predicted yields of
molecule-state $\overline{\Lambda n}$ and $\Lambda\Lambda$
as well as six-quark-state $\overline{\Lambda n}$ are higher
than the experimental upper-limits, the yield of six-quark-state
$\Lambda\Lambda$ could be lower than the upper-limits.
Therefore, the molecular or six-quark $\overline{\Lambda n}$
as well as molecular $\Lambda\Lambda$ are unlikely to be bound states.
However, the six-quark-state $\Lambda\Lambda$ could be a bound state
and cannot be excluded by the ALICE measurement.
If $\Lambda\Lambda$ is a multi-quark state, then according to the predicted
yield in the present work, we are likely at the edge of discovering it.

\begin{acknowledgments}
We are grateful to Jia-Lun Ping, Rui-Qin Wang and Zhong-Bao Yin for
helpful discussions. This work was supported in part by the Major
State Basic Research Development Program (973 Program) in China under
Contract Nos. 2015CB856904 and 2013CB834405, the NSFC under Grant
Nos. 11625521, 11275125 and 11135011, the ``Shu Guang" project supported by
Shanghai Municipal Education Commission and Shanghai Education
Development Foundation, the Program for Professor of Special
Appointment (Eastern Scholar) at Shanghai Institutions of Higher Learning,
and the Science and Technology Commission of Shanghai
Municipality (11DZ2260700).

\end{acknowledgments}

\end{document}